\documentclass{aa}  

\usepackage{graphicx}
\usepackage{txfonts}
\usepackage{lipsum}
\usepackage{subcaption}         
\usepackage{lscape}             
\usepackage{placeins}           
                                
\usepackage{color}	
\newcommand{\clr}{black}
 
\begin{document}

   \title{The atmosphere of K2-18 b}

   \subtitle{The role of hazes, clouds and photoelectrons}


	 \author{P. Lavvas\inst{1,2}
        \and R. Liu \inst{3} 
        \and G. Tinetti \inst{4}
        \and S. Paraskevaidou \inst{1}
        \and P. Drossart \inst{2}
        \and A. Coustenis \inst{5}
        }

   \institute{Laboratoire Environnements et Atmosphères Terrestres et Planétaires, Université Reims Champagne Ardenne, France\\
             \email{panayotis.lavvas@univ-reims.fr}
             \and Institut d'Astrophysique de Paris, UMR CNRS 7095, Paris, France
            \and University College London, London, UK 
            \and King's College London, London, UK
            \and Laboratoire d'Instrumentation et de Recherche en Astrophysique (LIRA), Observatoire de Paris, Université PSL, Sorbonne Université, Université Paris Cité, CY Cergy Paris Université, CNRS, 92190, Meudon, France
           }

   \date{Received September 30, 2025}
  \date{}

 
  \abstract
   {The atmospheric characterisation of temperate exoplanets is now becoming accessible with JWST, providing a critical connection between Solar System planets and the more commonly observed hot-Jupiters. K2-18 b, a temperate sub-Neptune orbiting an M dwarf, has emerged as a benchmark case following extensive JWST observations and ongoing debate regarding its atmospheric composition.}
   {We investigate the atmosphere of K2-18 b using a self-consistent forward model in order to constrain its metallicity, composition, and thermal structure, with particular emphasis on the role of disequilibrium chemistry, photochemical hazes and clouds. For the first time in this context, we also assess the impact of photoelectrons on the atmospheric chemistry of a temperate exoplanet.} 
   {We employ a one-{\color{\clr}dimensional} model that couples stellar energy deposition, disequilibrium gas-phase chemistry, and haze/cloud microphysics to generate physically consistent atmospheric scenarios. We explore a wide range of metallicities and intrinsic temperatures, evaluate haze and cloud formation, and compare the resulting transmission spectra with available JWST observations reduced using multiple independent pipelines.}
   {We demonstrate that a high metallicity (200-400$\times$solar) H$_2$-rich atmosphere consistently reproduces the observed transit spectra of K2-18 b, largely independent of the data reduction pipeline used. The atmospheric composition is strongly shaped by disequilibrium chemistry, with CH$_4$ dominating the spectrum alongside significant contributions from CO$_2$ and OCS, and a potential contribution from C$_2$H$_4$ at mid-infrared wavelengths. Photochemical hazes play a key role in shaping the thermal structure, producing a temperature minimum near the 10–100 mbar level that enables efficient condensation of H$_2$O and suppresses its gaseous abundance in the region probed by transit observations. Photoelectrons enhance the production of several disequilibrium species, particularly nitrogen-bearing molecules, although their direct impact on the current transmission spectra remains limited. Under sufficiently strong haze cooling, condensation of NH$_4$SH provides a natural explanation for the apparent absence of NH$_3$ in the observed spectra.}
   {Our results indicate that the JWST observations of K2-18 b are best explained by a hazy, high-metallicity sub-Neptune atmosphere shaped by disequilibrium chemistry. The combined effects of photochemical hazes and cloud formation are essential for interpreting the current K2-18 b observations. While uncertainties remain regarding haze optical properties, no additional molecular species beyond those considered here are required to reproduce the observed spectra.}


   \maketitle
   \nolinenumbers 

\section{Introduction}

The era of \it Hubble Space Telescope\rm ~and other space-borne and ground-based observatories has established and nurtured the field of exoplanetary research, identifying the vast variety of planetary environments present in our galaxy. With the dawn of the \it James Webb Space Telescope\rm ~era and the forthcoming observatories focused on exoplanetary research (e.g. ARIEL, ELT, HWO), the comprehension of the individual properties of each planetary case is excitingly improving, pushing the atmospheric characterisation boundaries towards the temperate regime (T$_{eq}$$\sim$300-500 K) or even below \citep{Encrenaz18,Encrenaz22}. Studying these atmospheres allows to reduce the gap between the planets of the solar system and the dominantly {\color{\clr}observed} hot-Jupiters and provides a key step towards the study of habitable exoplanets. 

\begin{figure*}[!h]
\centering
\includegraphics[scale=0.45]{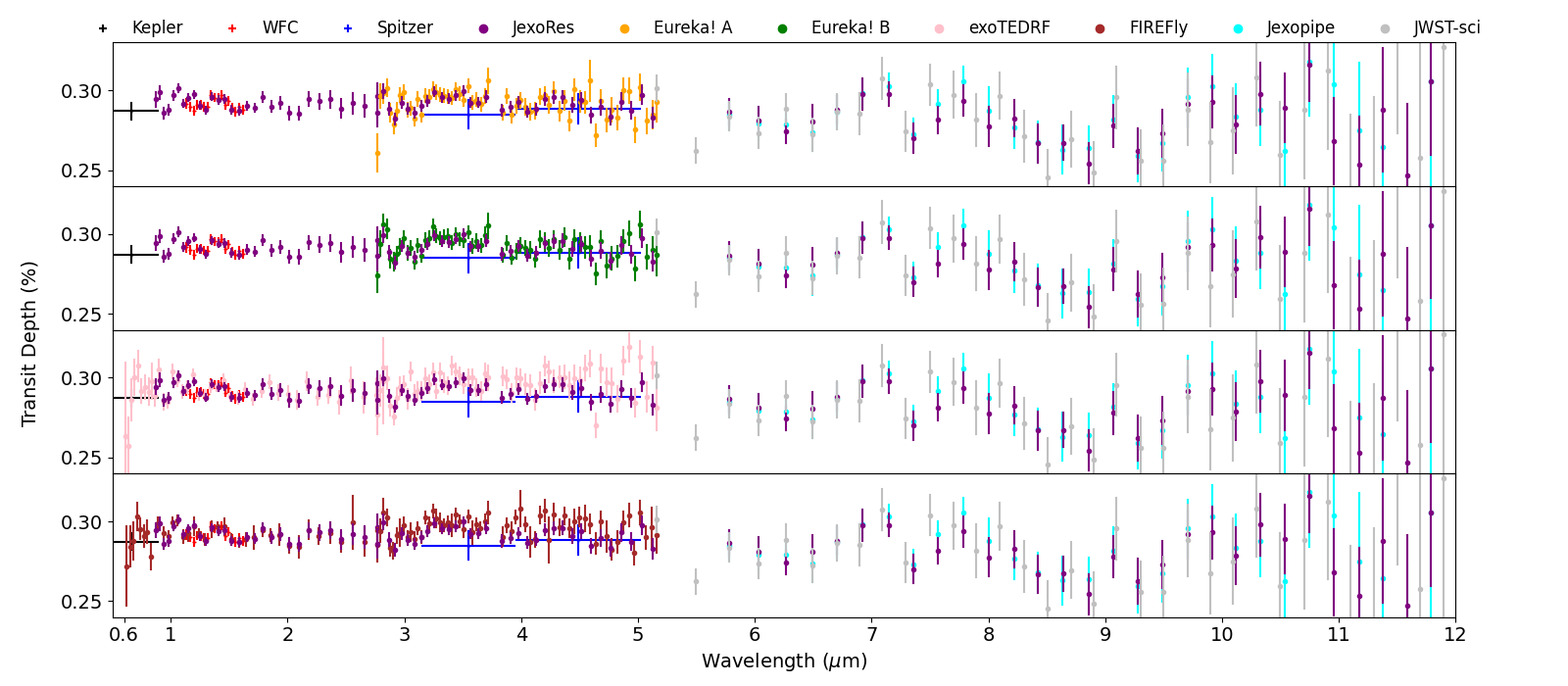}
\caption{Transit spectrum of K2-18 b based on the detection with Kepler \citep{Montet15} and characterisation with HST/WFC \citep{Benneke17}, Spitzer \citep{Benneke19}, JWST/NIRISS \&/NIRSpec \citep{Madhusudhan23, Schmidt25} and JWST/MIRI \citep{Madhusudhan25, Liu25}. Different panels compare different pipelines for the NIRISS and NIRSpec analyses.}\label{observations}
\end{figure*}

K2-18 b stands as a showcase of such planets and has stimulated strong debates in the field and a broad public interest. Following its initial discovery with Kepler \citep{Montet15} and radial velocity measurements \citep{Cloutier17, Cloutier19, Sarkis18}, the first attempts for atmospheric characterisation were performed with \it Hubble~\rm and \it Spitzer~\rm transit observations \citep{Benneke17,Benneke19} . The limited spectral range and resolution of these instruments sparked the first debate over the dominance of H$_2$O \citep{Tsiaras19} or CH$_4$ \citep{Bezard22} in the atmosphere of K2-18 b. Subsequent observations with JWST/NIRISS \& /NIRSpec \citep{Madhusudhan23} and JWST/MIRI \citep{Madhusudhan25} verified the presence of CH$_4$ and the lack of H$_2$O and NH$_3$ in the atmosphere of K2-18 b. Initial indications for the presence of CO$_2$ in the transit spectrum have been set in doubt when the observations were analysed with different pipelines \citep{Schmidt25}{\color{\clr}, but follow up observations from \cite{Hu25} appear to confirm the presence of CO2.} Suggestions for the presence of dimethylsuflide (DMS) and dimethyldisulfide (DMDS) that sparked a larger public interest due to the potential biological source of these species, are challenged from multiple other retrieval studies that allowed the reproduction of the observed spectra with other -simpler- molecules, which nevertheless constitute important advancements in the characterisation of exoplanetary atmospheres \citep{Welbanks25, Liu25, Fernandez25, Stevenson25}. In Fig.~\ref{observations} we compare the available transit spectra obtained through different pipelines used for the reduction of the JWST measurements and contrast them to older observations. Although differences among the pipelines are present and affect the reported retrievals, the transit spectra present an overall consistent picture. However, what is striking among the different observations is the large amplitude of transit depths probed with MIRI compared to those from NIRISS and NIRSpec, suggesting a larger altitude (pressure) range probed at MIRI wavelengths. However, the MIRI observations have a higher uncertainty than those at shorter wavelengths and their potential to constrain the atmospheric composition is questioned \citep{Luque25,Stevenson25,Taylor25}.

The K2-18 b observations motivated the exploration of different scenarios regarding the nature and properties of its atmosphere and interior. Initial estimates for the density of K2-18 b suggested a super-Earth type planet \citep[$\rho$$\sim$4 g/cm$^3$,][]{Sarkis18}, but were progressively refined towards a sub-Neptune \citep[$\rho$$\sim$2.3 g/cm$^3$,][]{Howard25}. Characterising the atmospheric structure and composition of this planet can provide important information for the mechanisms at play at this boundary region between small planets with potentially secondary atmospheres or those preserving a significant part of their H$_2$-rich primordial envelope \citep{Fulton17, Owen18}. For this reason, scenarios of different type planets were put forward and evaluated against the observations. Based on its density and orbital distance, \cite{Madhusudhan21} proposed K2-18 b to be an ocean covered planet with a thin H$_2$ atmosphere {\color{\clr}\citep[Hycean planet, see also][]{Madhusudhan20,Nixon21}}, while in an attempt to explain the apparent lack of NH$_3$ in the observed transit spectra, \cite{Shorttle24} suggested the presence of a magma ocean that would allow the solubility of nitrogen species at reducing conditions, although the conditions and efficiency for solubility depend on multiple parameters \citep{Rigby24}. In parallel, scenarios based on H$_2$-rich atmospheres were explored \citep{Hu21,Wogan24} and found consistent with most of the observations. Irrespective of the planet type the lack of H$_2$O in the atmosphere of K2-18 b is attributed to condensation giving rise to H$_2$O clouds \citep{Huang24}, while improved constraints on the CO$_2$/CO ratio could potentially aid in revealing the presence or not of a surface \citep{Yu21,Luu24}. An evaluation of the planetary albedo could further constrain the hycean scenario because a high value is required to keep the surface temperature low-enough to sustain water condensation \citep{Leconte24} and the presence of clouds/hazes would thus be necessary. Initial estimates for the required albedo ($\sim$50-60$\%$) were significantly higher than the albedo estimated from the transit observations \citep[$\sim$20$\%$,][]{Jordan25}, while more recent evaluations based on an improved treatment of the atmospheric scattering still suggest values as low as 27$\%$ for an 1 bar atmosphere \citep{Barrier25}, thus advocating for the H$_2$-rich atmosphere scenario. In parallel, anticipations for the presence of hazes in this atmosphere are becoming prominent \citep{Liu25,Jaziri25}, while retrievals based on non-equilibrium chemistry \citep{Jaziri25b}, further support the dominance of such processes in the atmosphere of K2-18b.

In view of these recent developments, the current study presents results from 1D forward models that evaluate in detail the anticipated properties of gases, photochemical hazes $\&$ clouds and their impact on the thermal structure and the planetary albedo. In addition we explore novel (to our knowledge) contributions to the exoplanet atmospheric disequilibrium chemistry. We emphasise that our approach is not a retrieval and does not aim to identify a unique best-fit solution. Instead, we explore whether physically self-consistent atmospheric scenarios can reproduce the key features of the observed transmission spectra and identify which processes are required by the data.

In the following section we provide a quick overview of the methods applied and in section 3 we present our nominal results based on which we evaluate, through comparison with the transit observations, the optimal atmospheric metallicity for K2-18 b when treated as a sub-Neptune. In the following sections we investigate the role of photoelectrons on the atmospheric chemistry (Section 4), we further evaluate possible processes that could partake in the interpretation of the MIRI observations (Section 5), and explore a possible mechanisms to explain the apparent lack of NH$_3$ in the observed transit spectrum (Section 6). In section 7, we discuss additional aspects that could contribute in the interpretation of the observations, before our final conclusions.

\section{Methods}

The latest evaluations for the mass density of K2-18 b provide values between  $\rho$=2.67$^{+0.52}_{-0.47}$ g/cm$^{-3}$ \citep{Benneke19} and $\rho$=2.28$^{+0.63}_{-0.51}$ g/cm$^{-3}$ \citep{Howard25}, consistent with a sub-Neptune planet. Therefore, we explore its possible atmospheric composition by considering different metallicity scenarios using a forward model that couples among radiative transfer, disequilibrium chemistry and haze/cloud microphysics \citep{Arfaux22} and benefits from the experience gained from past studies of Titan's and Pluto's atmosphere in the solar system \citep[e.g.][]{Vuitton19, Lavvas21b}.

\begin{figure}
\centering
\includegraphics[scale=0.6]{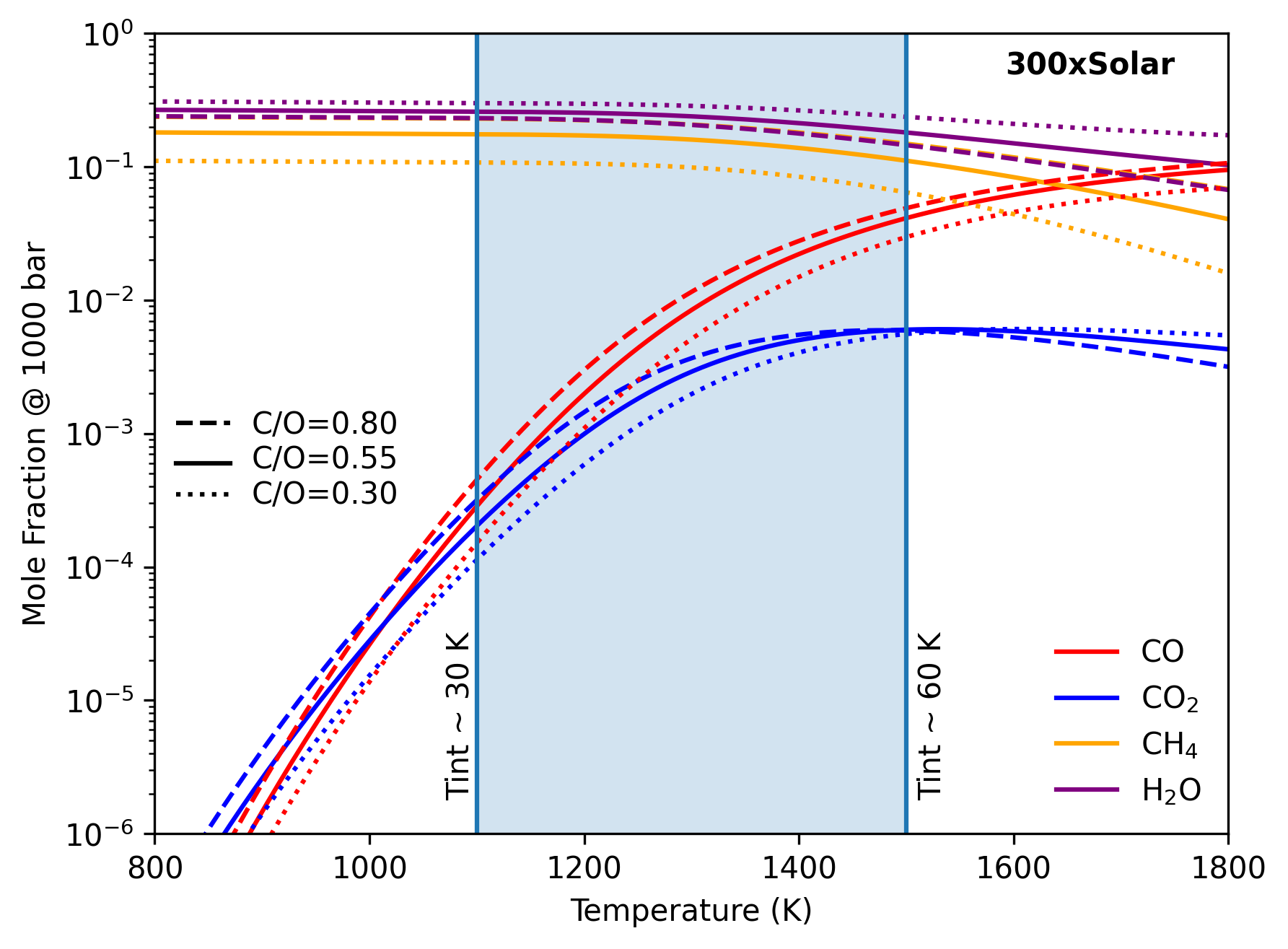}
\caption{Equilibrium abundances of H$_2$O, CO, CO$_2$ and CH$_4$ for a 300x solar metallicity case across different temperatures at 1000 bar. Different style lines show the implications of changing the C/O ratio.}\label{boundary}
\end{figure}

We simulate the atmospheric structure at pressures between 10$^3$ bar and 10$^{-10}$ {bar} assuming complete energy redistribution of the incoming instellation. At the lower boundary we consider that the composition is equal to the thermochemical equilibrium at the calculated temperature. Equilibrium simulations based on the GGchem model \citep{Woitke18} demonstrate that at temperatures between 1000 K and 1500 K expected at the lower boundary, the CO and CO$_2$ abundances change by more than a factor of 10 (see e.g. Fig.~\ref{boundary} for the case of 300x solar metallicity). The relative abundance of C and O (the C/O ratio) also affects the resulting abundance of the main carbon bearing species but to a smaller degree than the dependence on temperature (Fig.~\ref{boundary}). Therefore, the main parameter affecting the boundary conditions of our simulations is the intrinsic temperature (T$_{int}$) assumed, a conclusion consistent with previous studies \citep[e.g.][]{Wogan24}. We thus consider a solar C/O=0.55 \citep{Lodders19} and evaluate two cases of  T$_{int}$ = 30 K and T$_{int}$ = 60 K, based on estimates and observational constraints on this parameter \citep{Blain21,Wogan24}, to investigate how this parameter affects the resulting atmospheric composition and the interpretation of the observations. 

Our chemical composition simulations are performed through a network of chemical reactions considering H/C/N/O/S composition \citep[see][and references therein]{Arfaux24}. The chemical composition is disturbed from the equilibrium established deep in the atmosphere due to the atmospheric mixing and the stellar flux. We describe the atmospheric mixing through a K$_{ZZ}$ profile that is based on results from general circulation models \citep{Charnay15a}. Although the mixing can depend on the assumed atmospheric metallicity, we use a fixed profile across the different metallicity cases that allows to separate the effects of mixing from other processes (e.g. thermal structure and viscosity) in our evaluations.

K2-18 is a M2.5V star in the constellation of Leo at a distance of 38 parsecs. Observations of its UV spectral flux are limited to HST/STIS measurements with the G140M (PI:Ehrenreich, ID:14221) and G230L $\&$ G430L grisms (PI:Youngblood, ID:16701), and more recently observations with XMM-Newton \citep{Rukdee25}. Due to the lack of a full instellation spectrum the flux of K2-18 is typically estimated in comparison with that of GJ 176, a similar type M star for which a full spectrum is available from the MUSCLES database \citep{Brown23}. Comparison of the observational constraints with the estimated spectrum reveals similar fluxes at visible and NUV wavelengths but with differences increasing towards shorter wavelengths, although the low signal to noise ratios in the measured K2-18 FUV fluxes do not allow for clear conclusions. Similarly, despite its large uncertainty, the measured Ly-$\alpha$ flux is $\sim$ 4$\times$ larger than that of GJ 176, while a similar factor applies for the expected EUV flux \citep{Santos20}. These differences between the stellar fluxes at high energies are probably relevant to the different ages of the two stars, with K2-18 \citep[$\sim$2-3 Gyr,][]{Guinan19,Sairam25} being markedly younger than GJ 176  \citep[8.8 Gyr,][]{Brown23}. In our calculations we adopted as nominal stellar spectrum of K2-18 the spectrum of GJ 176, but we explore the implications of an increased UV flux.

Our simulations also include the opacity of photochemical hazes evaluated from the microphysics core of the model. The presence of such particles has multiple implications for the atmospheric properties, affecting the thermal structure, the albedo and the photochemistry in the simulated planetary atmospheres \citep[][and references therein]{Lavvas24}. We estimate the haze mass flux from the photolysis of typical precursors considered as sources for the formation of photochemical hazes \citep{Lavvas19}.  For all metallicity cases we consider two scenarios for the photochemical haze mass flux, one with $\Phi_H$=10$^{-12}$ g cm$^{-2}$s$^{-1}$ based on our photochemistry simulations (see below) and a second scenario of a lower photochemical haze mass flux of $\Phi_H$=10$^{-13}$ g cm$^{-2}$s$^{-1}$ for which hazes would have a negligible impact on the atmospheric thermal structure. 
The photochemical haze production is simulated as 1 nm radius particles formed at 1 $\mu$bar and we follow their microphysical evolution as the particles settle and are mixed deeper in the atmosphere, while growing in size due to their random collisions. As {\color{\clr}a} nominal case we assume spherical particles and evaluate their optical properties considering the refractive index from laboratory experiments at similar exoplanetary conditions \citep{He24}, but we discuss further additional scenarios for their optical properties.

\begin{figure*}[!h]
\centering
\includegraphics[scale=0.5]{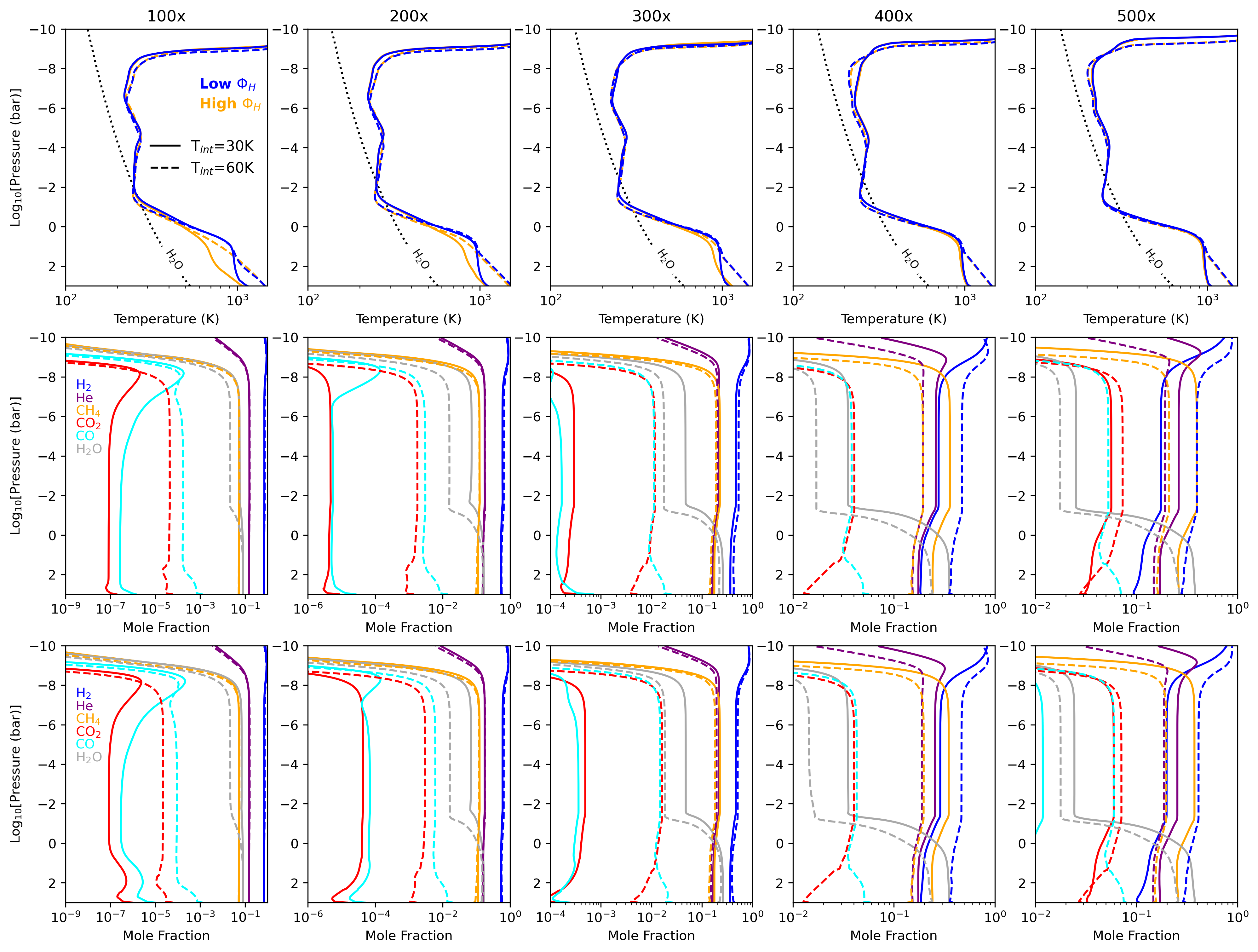}
\caption{Thermal structure (top row) for the different metallicity cases considered. Blue (orange) lines correspond to the low (high) photochemical haze mass flux scenario, while solid and dashed lines correspond to the T$_{int}$ = 30 K and 60 K cases, respectively. The middle (high $\Phi_H$)  and lower (low $\Phi_H$) panels present the corresponding disequilibrium composition for the high (solid) and low (dashed) T$_{int}$ cases, respectively.}\label{pTcomp}
\end{figure*}

Our simulated temperature profiles shown below reveal that between 10$^{-2}$ and 10$^{-3}$ bar the temperature drops sufficiently to allow for water condensation, in agreement with previous anticipations \citep{Benneke19}. We include this loss in the simulated H$_2$O profiles by taking into account the formation of clouds in the microphysics calculations considering that haze particles act as nucleation sites for the condensation of water vapour. For the nucleation and subsequent condensation processes we considered H$_2$O saturation pressure from \cite{Fray09} and surface tension from the Dortmund Data Bank\footnote{http://ddbonline.ddbst.com}. We assume a small contact angle ($\mu$=0.995) between the haze particles and water condensate and consider liquid water formation. The interaction of water with possible exoplanetary photochemical hazes is rather unconstrained and the small contact angle assumption implies a rapid formation of clouds when condensation conditions are met.

We thus evaluate the atmospheric thermal structure, the disequilibrium chemical composition and the particle (haze $\&$ cloud) size distributions and use these results in a transit simulator to evaluate the possibility of each scenario with respect to the available transit observations. {\color{\clr}The transit simulator we use is described in previous studies \citep[see][and references there in]{Arfaux22} where information for the opacity sources considered can be found. We use the correlated-k method at 1 cm$^{-1}$ resolution for the simulation of the transit spectra. We finally note that we do not attempt to search for the optimal parameters that would allow the best fit of the observed spectra (i.e. an inversion). Instead we explore a range of well defined scenarios with our forward model and explore how each scenario compares with the observations and how this informs on the possible physical and chemical processes in the atmosphere.}

\section{Nominal results}

\subsection{Thermal structure \& chemical composition}
For the low haze mass flux and T$_{int}$ = 30 K the resulting temperature profiles for metallicities between 100$\times$ and 500$\times$ solar metallicity (Fig.~\ref{pTcomp}, top row) demonstrate a hot deep atmosphere followed by a temperature drop near $\sim$1 bar and a quasi-isothermal thermal structure at lower pressures before the onset of the thermosphere at pressures {\color{\clr}lower than} 10$^{-8}$ bar.  As metallicity increases the opacity of the atmosphere rises and photon penetration in the atmosphere is constrained to progressively lower pressures resulting in a slight increase of the transition region between the hot deep atmosphere and the cooler upper atmosphere. Increasing the T$_{int}$ assumed in the simulations from 30 K to 60 K, strongly affects the deep atmosphere raising the temperature at the lower boundary by $\sim$400 K. The simulated temperature profiles for the two T$_{int}$ cases merge at pressure{\color{\clr}s lower than} $\sim$1 bar. Nevertheless, the temperature differences at the lower boundary impose significant modifications on the gas phase composition (Fig.~\ref{pTcomp}, bottom row), which propagate to the upper atmosphere as the composition is perturbed from thermochemical equilibrium before reaching the 1 bar level. Under all metallicity cases, CH$_4$ is the dominant spectroscopic abundance followed by H$_2$O whose abundance drastically reduces near the cold trap of each studied case and remains constant at lower pressures to the mole fraction imposed by the temperature minimum of each atmosphere. CO$_2$ reveals the expected dependence to the T$_{int}$ value with higher mole fractions propagating to probed atmosphere for increasing T$_{int}$. We discuss further other disequilibrium chemistry products in the following section.

\begin{figure*}
\centering
\includegraphics[scale=0.45]{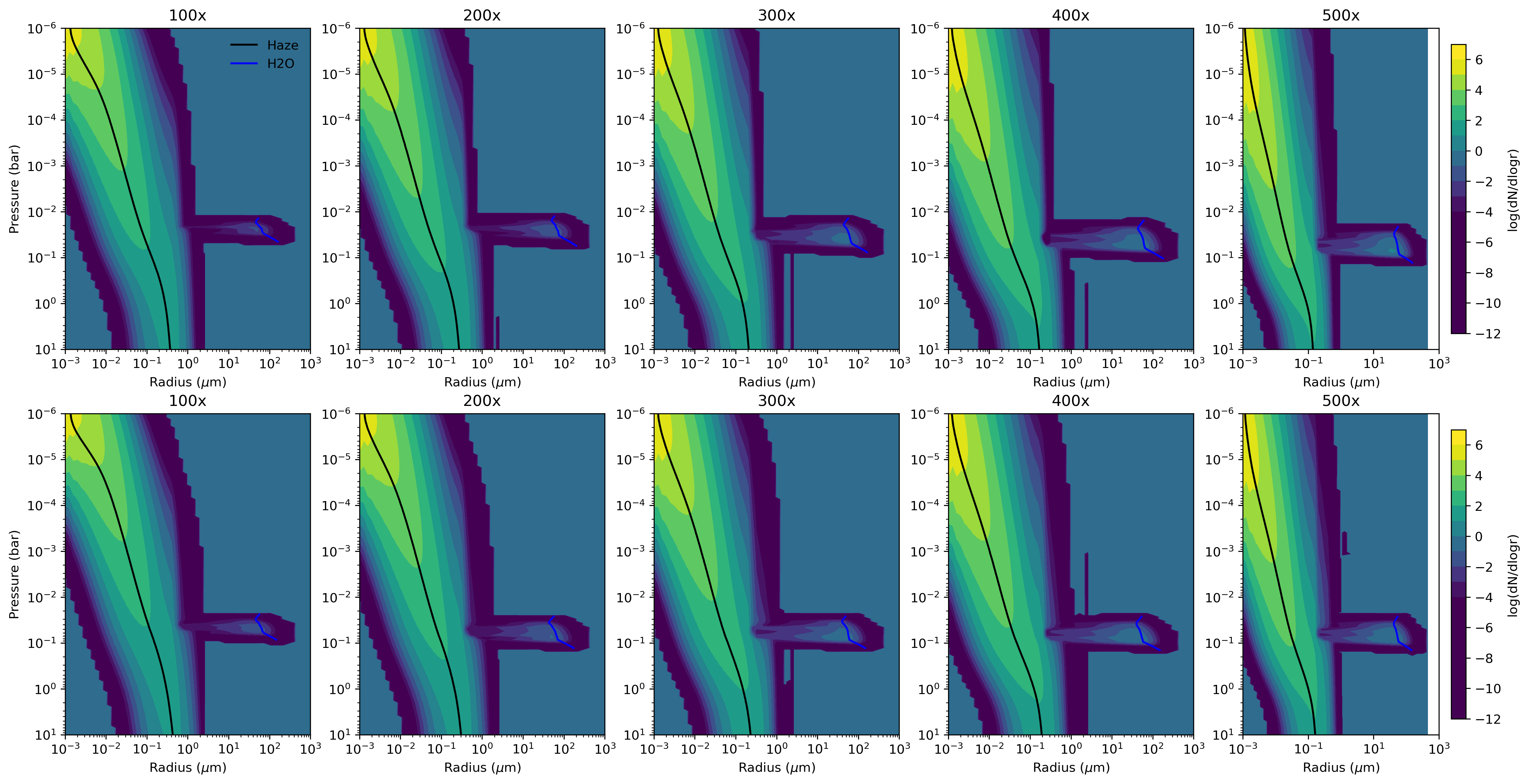}
\caption{Particle size distributions including both photochemical hazes and H$_2$O clouds for different metallicity cases. The top row corresponds to T$_{int}$=30K and the bottom row to T$_{int}$=60K.}\label{particles}
\end{figure*}

\begin{figure*}
\centering
\includegraphics[scale=0.6]{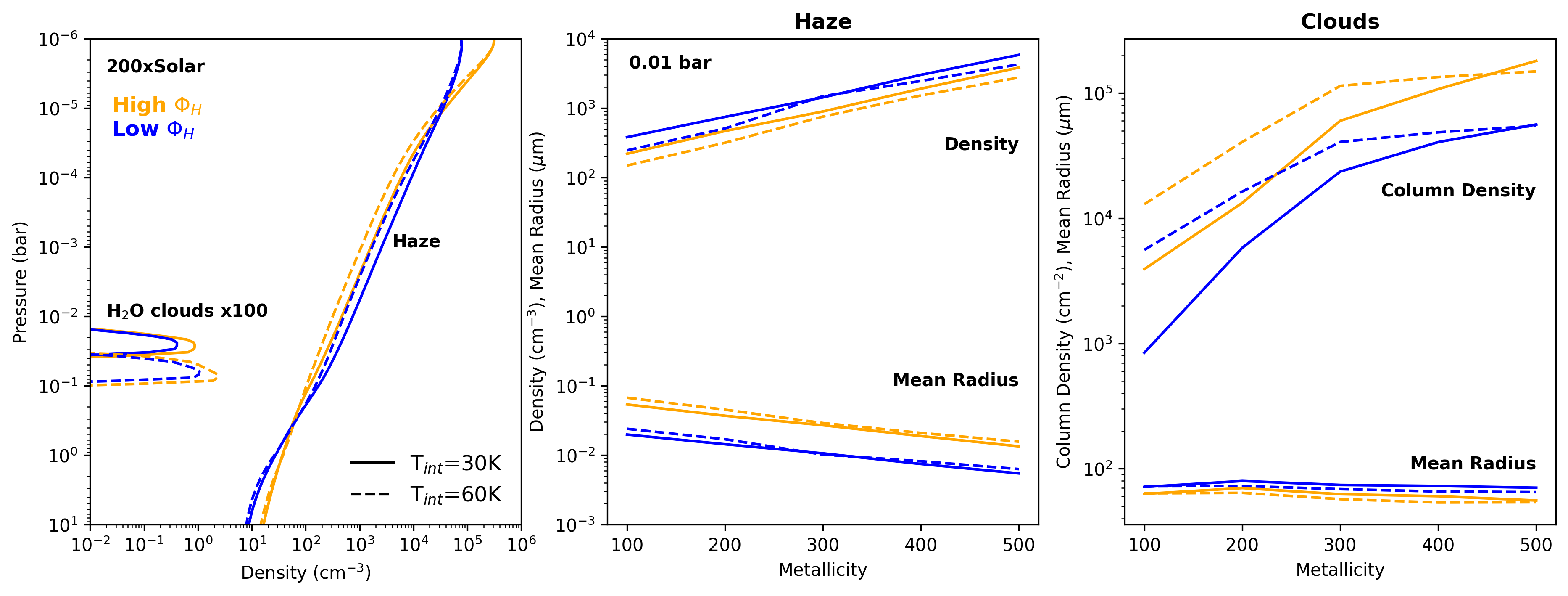}
\caption{Mean particle properties for hazes and H$_2$O clouds. The left panel present the mean density profiles for the 200x solar metallicity case for the different cases of T$_{int}$ (line styles) and $\Phi_H$ explored (line colours). The middle and right panels present the mean particle properties across different metallicities for hazes and clouds, respectively. For the haze we present results at 0.01 bar, while for the clouds the integrated mean values over the narrow altitude range of their existence.}\label{particles_metal}
\end{figure*}

Starting with the high $\Phi_H$ case, particles are abundant enough to scatter a significant part of the incoming radiation. As a result, the temperature in the deeper atmosphere is reduced relative to the low  $\Phi_H$ case, with the differences being more prominent for the low T$_{int}$ condition. Moreover, the impact of hazes on the thermal structure is progressively less significant with increasing metallicity as the gaseous opacity increases and the added opacity by the hazes is relatively less important. Practically, for the 400$\times$ and 500$\times$ solar metallicity cases, the temperature profiles are identical for the two $\Phi_H$ cases explored (Fig.~\ref{pTcomp}). The differences in the gaseous composition due to the higher haze mass flux are thus prominent at lower metallicity cases, however they remain smaller than the differences from the assumption on T$_{int}$.

We note that the presented temperature profiles do not include the effect of the cloud opacity. This is done because first its effect is small, resulting to slightly cooler temperature above the cloud top (by less than 10 K) and warmer below, and second because cloud formation is sensitive to temperature variations and including the opacity of clouds results in numerical oscillations due to the cloud-temperature feedback that limit the overall convergence of the simulations. Thus, we remove this opacity contribution from the thermal structure but we keep in mind its impact in the following. 

\subsection{Haze \& cloud properties}

\begin{figure*}[!h]
\centering
\includegraphics[scale=0.45]{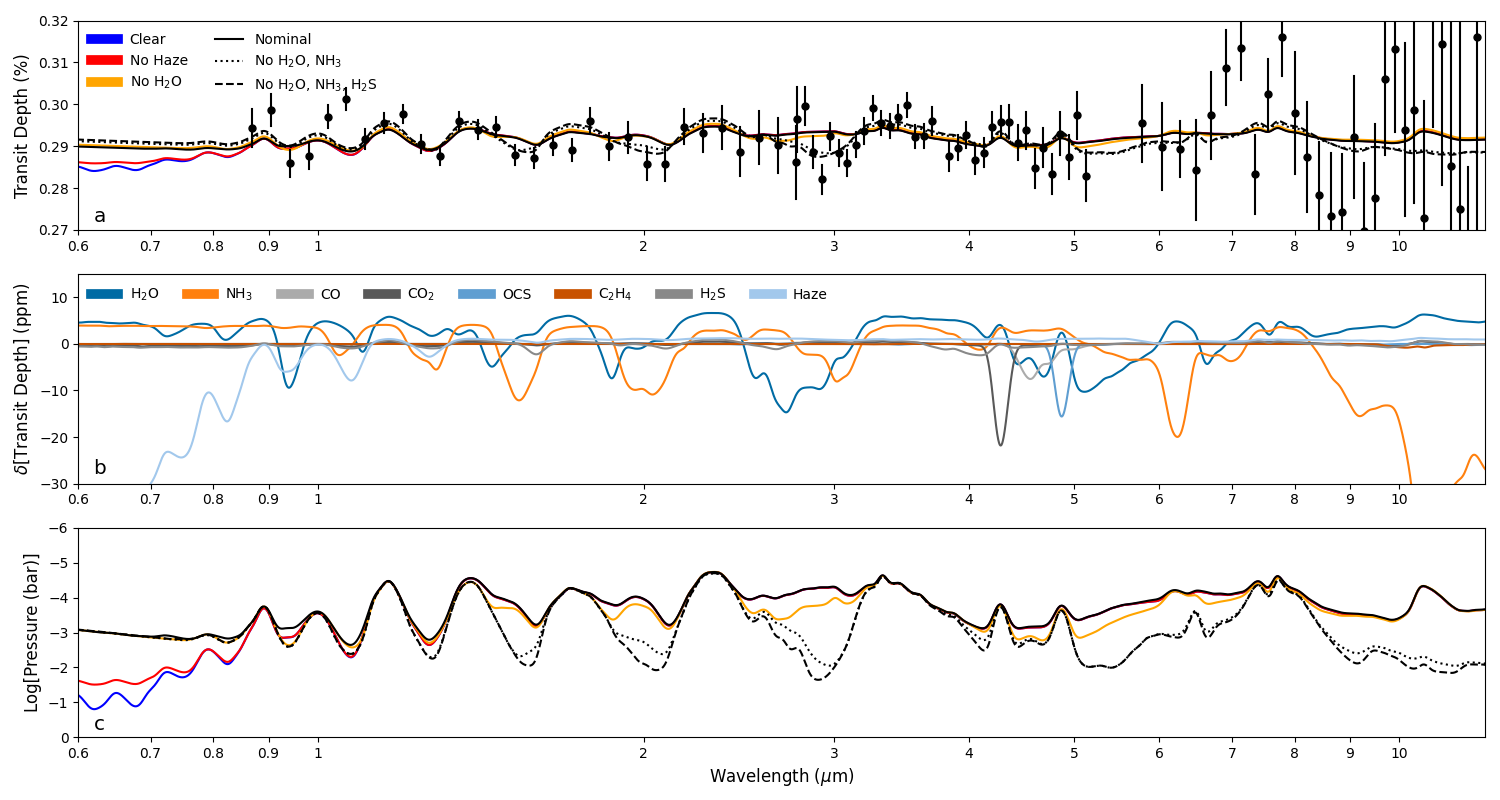}
\caption{a: Simulated transit spectra of K2-18b at R=100 (lines) compared to the observed spectrum from the JExoRES pipeline \citep{Madhusudhan23,Madhusudhan25}. The black line corresponds to the 300$\times$ solar metallicity scenario with high $\Phi_H$ and T$_{int}$=60K. b: Differential transits for individual components removed from the transit simulations. c: Pressures probed for the different scenarios explored in panel a.}\label{transit}
\end{figure*}

We can estimate the photochemical haze production by evaluating the photolysis rates of the main photochemical species in the upper atmosphere of K2-18 b (above $\sim$10 $\mu$bar). Our simulations reveal that across different metallicities the mass flux generated by the photolysis of such precursors is significant and varies between (0.5-1.5)$\times$10$^{-11}$ g cm$^{-2}$s$^{-1}$ (see discussion section for further details). However, not all of the generated mass flux will end in the formation of the photochemical hazes. Thus, our nominal photochemical haze mass flux of $\Phi_H$=10$^{-12}$ g cm$^{-2}$s$^{-1}$ corresponds to haze formation yields of 7-20$\%$, i.e. a conservative formation based on estimates from Titan's \citep{Vuitton25} or Jupiter's \citep{Wong00} atmospheres, while the low $\Phi_H$ scenario considered corresponds to a factor of 10 reduction in this yield. 

The simulated size distributions of photochemical hazes reveal a monotonic growth of the haze particles from their production region at 1~$\mu$bar to the deep atmosphere (Fig.~\ref{particles}), reaching average radii of $\sim$0.1 $\mu$m near 0.1 bar, at the expense of their number density. The maximum mean size reached for the haze particles depends on the haze mass flux and the metallicity assumptions. A higher mass flux results {\color{\clr}in} larger particles, while metallicity has multiple implications on the particle growth. Changes {\color{\clr}to} the thermal structure control the rate of coagulation, while changes in the atmospheric composition control the atmospheric viscosity that imparts on the particles settling velocities. Comparing the resulting mean particle properties at 0.01 bar across the metallicity cases considered (Fig.~\ref{particles_metal}) we see that increasing the metallicity results in smaller mean particle radii in tandem with an increase of their corresponding number density.

Our simulations reveal that water clouds grow rapidly to radii between 50 and 80 $\mu$m but their number densities remain small compared to those of the haze and reach maximum values near 10$^{-2}$ cm$^{-3}$ (Fig.~\ref{particles_metal}). As expected from the simulated temperature profiles (Fig.~\ref{pTcomp}), the H$_2$O cloud layer becomes progressively larger with increasing metallicity as a larger region of the atmosphere allows for the water condensation. The high $\Phi_H$ case allows for more cloud particles to form, with a corresponding decrease in the mean particle radius, while changing the T$_{int}$ affects the location of the cloud layer, with the high T$_{int}$ cases resulting to clouds slightly deeper in the atmosphere. Metallicity effects have a small impact on the particle size, since the resulting cloud particles are large and their abundance is limited by their settling. As the cloud particle number density is narrow we calculated the cloud particle column density to evaluate the variation of cloud particles with metallicity (Fig.~\ref{particles_metal}). The results reveal an increasing column density with increasing metallicity that reflects the larger cloud layer thickness.

\begin{figure*}
\centering
\includegraphics[scale=0.42]{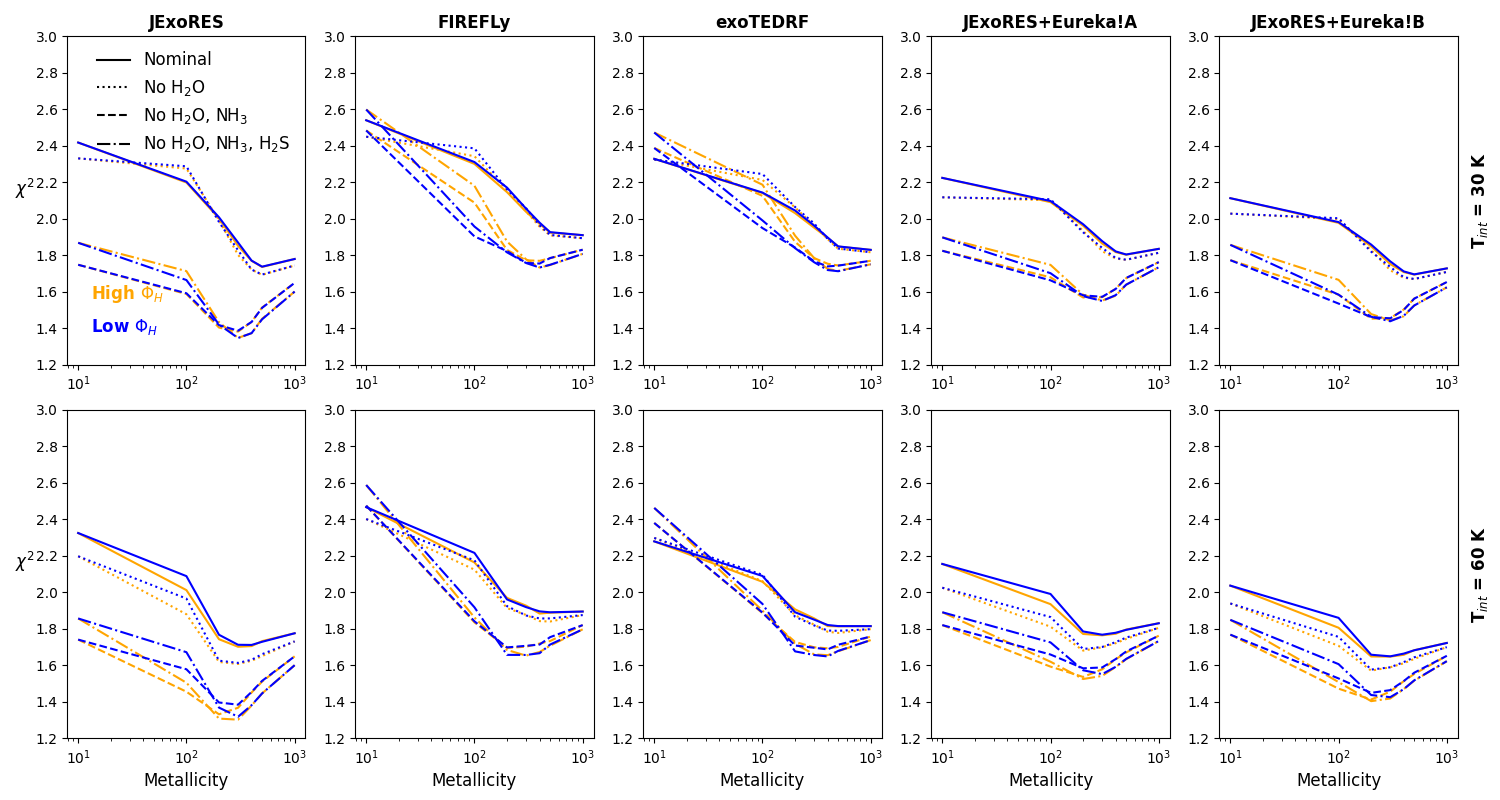}
\caption{Reduced $\chi^2$ distributions across metallicity with various JWST pipelines reported in the literature and for the different scenarios of the forward model regarding the haze mass flux and T$_{int}$.}\label{chi2}
\end{figure*}

\subsection{Transit spectra}

The simulated atmospheric structure, disequilibrium gas composition and haze/cloud size distributions are then used in a transit simulator to evaluate the relevance of each studied case to the observational constraints. To understand better the role of each opacity contributor, Fig.~\ref{transit}a presents the transit spectrum for the 200x solar metallicity, the high $\Phi_H$ and T$_{int}$ = 60 K case, along with simulations with individual opacity components removed. For visual clarity in the description of the qualitative behaviour of the simulated spectra we compare with the observations obtained with the JExoRes pipeline \citep{Madhusudhan23, Madhusudhan25}. The NIRSpec spectrum is shifted by -41 ppm as suggested \citep{Madhusudhan23}, while we shifted the MIRI spectrum {\color{\clr}(-105 ppm)} so that its mean transit depth is the same as the mean transit depth from the NIRISS and NIRSpec transits. The simulated transit spectra are pressure referenced to the NIRISS observations.   Our nominal case (black solid line) including haze, H$_2$O clouds and disequilibrium gases provides a transit  spectrum that {\color{\clr}follows the general behaviour of the NIRISS and NIRSpec observations in terms of the amplitude of the spectrum and the locations of the main spectral features observed, although regions where improvements could be sought for are present. On the contrary,} the amplitude of the simulated spectrum is significantly smaller than that of the MIRI observations, which however are characterised by larger uncertainty than the other JWST observations. At short wavelengths ($\lambda$$\lesssim$1 $\mu$m), haze dominates the atmospheric opacity and controls the simulated transit depth, while clouds that affect the atmospheric opacity at {\color{\clr}higher pressures} (Fig.~\ref{transit}c) have a negligible impact on the transit spectrum (see red line for no haze included) that is traceable only when comparing with a clear atmosphere (only gases) transit spectrum (blue line).

The differential transit depths (Fig.~\ref{transit}b) reveal the difference in the transit spectrum when comparing the nominal case spectrum with a simulated spectrum having one component removed. Note that removing a component can affect both the simulated opacity and mean molecular weight of the atmosphere. Moreover, such a modification can adjust also the pressure referencing of the simulated atmosphere. The differential transit spectra reveal that although H$_2$O condensation significantly reduces its gaseous abundance (see Fig.~\ref{pTcomp}) in the region of the atmosphere probed in transit (Fig.~\ref{transit}c), the residual water vapour still has a contribution to the transit spectra. In addition, opacity contributions from NH$_3$, CO$_2$ and OCS dominate the spectrum beyond the contribution of CH$_4$. Disequilibrium chemistry has an important role in defining the abundances of the gaseous species in the simulated atmosphere. Our simulated abundances of the species affecting the transit spectrum are orders of magnitude different than those anticipated by equilibrium calculations at the same temperature. We further note that C$_2$H$_4$ is at the limit of affecting the spectrum near 10 $\mu$m, indicating that if its abundance was slightly higher it would have a signature in the transit, as suggested by recent inversion results \citep{Liu25}. 

We now proceed to a quantitative comparison of our simulations with the observed transit spectra. As the MIRI observations reveal a larger uncertainty, we focus on the NIRISS (2$^{nd}$ order) and NIRSpec observations for this evaluation. Given the slightly different results on the observed spectra derived from the various pipelines (Fig.~\ref{observations}), we compare separately with each reduction in Fig.~\ref{chi2} in terms of the reduced $\chi^2$ of each studied case. {\color{\clr}We consider the spectral resolution provided in the published reductions (as shown in Fig.~\ref{observations}). The Eureka! reductions consider only the NIRSpec observations, thus we combine them with other pipeline reductions of NIRISS observations to get the full transit spectrum.} For a more global representations, the $\chi^2$ results also include simulations for 10$\times$ and 1000$\times$ solar metallicity cases that allow to probe the phase space of solutions away from the main region of interest that will be identified below. 

Focusing on the the nominal model cases (solid lines in Fig.~\ref{chi2}) that correspond to the full composition calculated by the forward model, the $\chi^2$ results demonstrate that atmospheres with metallicities greater than 100$\times$ solar are in better agreement with the observations, in agreement with the latest studies \citep[e.g.][]{Wogan24,Jaziri25}. For the JExoRES {\color{\clr}transit spectrum and the spectra from the two Eureka! reductions (when combined with the JExoRES NIRSISS spectrum)} the $\chi^2$ results present a local minimum between 300$\times$ and 500$\times$ solar metallicities, while for the FIREFLy and exoTEDRF spectra the $\chi^2$ curves present a plateau above 500$\times$ solar metallicity. Combining the Eureka! spectra with the FIREFLy or exoTEDRF NIRISS results does not change qualitatively the picture, although results in lower $\chi^2$ values and the minima become more shallow. Apart from metallicity, T$_{int}$ also affects the shape of the $\chi^2$ curves, with the T$_{int}$ = 60 K cases providing overall better fits and shifting the metallicity inferences to lower values than the corresponding 30 K cases. The improved fits with the higher T$_{int}$ are due to the stronger contribution of CO$_2$ in the atmospheric opacity that results into an observable signature in the simulated transit spectra near 4 $\mu$m consistent with most pipeline results. On the contrary, the haze mass flux does not appear to have a major impact on the quality of the fits for the nominal cases, particularly in the range of metallicities that are more consistent with the observed spectra. Although the impact of the haze is evident in the simulated spectrum, this result occurs because of the limited spectral range over which haze effects can be contrasted from the observations.

To explore further the differences between the inversion results and the forward models, we evaluated two additional scenarios. For the first we completely removed H$_2$O from the simulated atmosphere of the transit spectra. The new spectra under this scenario provide a slightly reduced $\chi^2$ values for most simulated cases and particularly those with T$_{int}$=60 K (dotted lines in Fig.~\ref{chi2}), indicating that the observations are consistent with a slightly cooler atmospheric temperature at the region of the H$_2$O cloud formation. Moreover, for the high T$_{int}$ case, the derived metallicity is reduced relative to the nominal results. This occurs because the removed H$_2$O opacity increases the structure in the simulated spectra (see orange line in Fig.~\ref{transit}a) in a manner that is consistent with the observations. The same effect is further magnified on the second scenario where in addition to H$_2$O we remove NH$_3$ (see dashed lines in Figs.~\ref{transit}$\&$\ref{chi2}). Under these conditions, the reduced $\chi^2$ is significantly reduced for all datasets and all simulated cases provide a clear minimum for the metallicity in the range of 200$\times$ to 500$\times$ solar, which can be further constrained to 200-400$\times$solar for the high T$_{int}$ cases that fit better the observed spectra. {\color{\clr} We also explored the case where we additionally remove H$_2$S that has a minor contribution in the transit spectrum at the throughs near 2 and 3 $\mu$m (compare dotted and dashed lines in Fig.~\ref{transit}), but this case does not modify the above metallicity picture for the atmosphere of K2-18 b (see dash-dotted lines in Fig.~\ref{chi2}).}

\begin{figure*}
\centering
\includegraphics[scale=0.48]{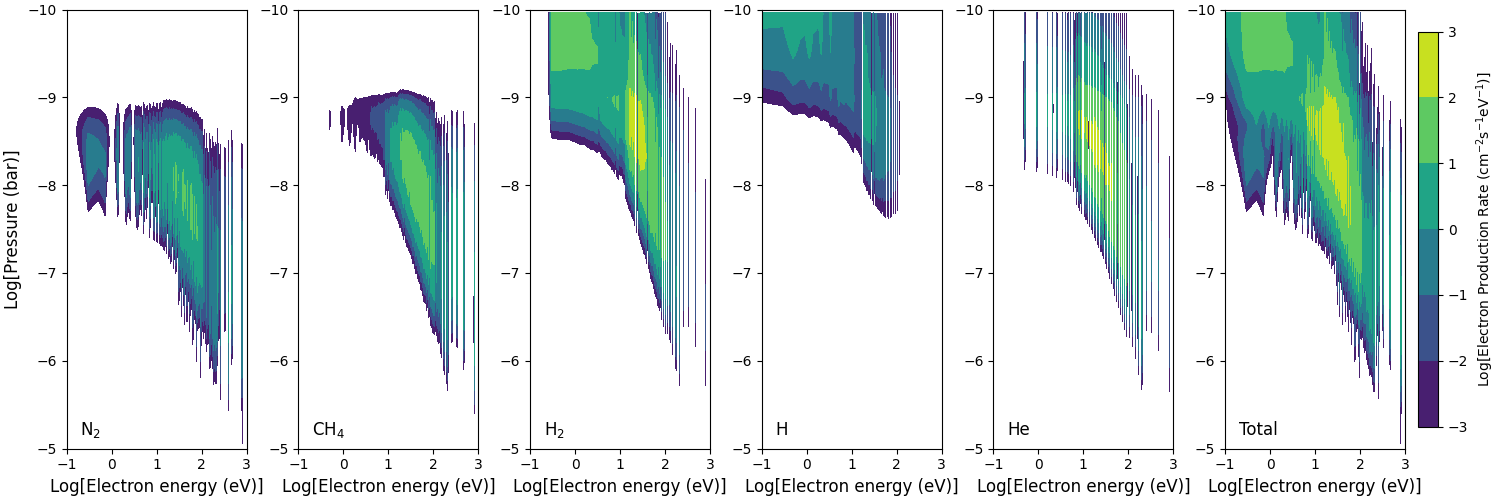}
\caption{Photoelectron production rates across energy from individual species and the cumulative production rate under nominal stellar flux conditions.}\label{photoelectrons}
\end{figure*}

As a conclusion of the above results, our forward simulations demonstrate that a high metallicity (200-400$\times$ solar) sub-Neptune atmosphere is consistent with the observed transit spectrum of K2-18 b, irrespective of the JWST pipeline used. Disequilibrium-driven chemistry products CO$_2$, OCS {\color{\clr}can have} a significant contribution to the observed spectrum along with the spectroscopically dominant CH$_4$. Photochemical hazes, if present, have a much more dominating role on the transit relative to H$_2$O water clouds, although the latter are critical for explaining the lack of H$_2$O on the observed spectrum. However, additional investigations are required to explain the need for an enhanced loss of H$_2$O (more than the loss anticipated from our cloud simulations) and the lack of NH$_3$ in the transit spectrum. Moreover, the stronger amplitude of the MIRI observations relative to those with NIRISS and NIRSpec and the possible contribution of C$_2$H$_4$ near 10 $\mu$m motivate further exploration of the processes affecting this part of the observed spectrum. We thus limit the following investigations to the case of 200$\times$solar metallicity with T$_{int}$ = 60 K, and investigate various processes that could aid in further interpreting the observational constraints. 

\section{Enhanced disequilibrium chemistry}

Typical disequilibrium chemistry simulations for exoplanet atmospheres consider only the role of high-energy photons from the star in the atmospheric photochemistry. However, UV photons able to ionise the main atmospheric gas species invoke the release of energetic electrons (photoelectrons) that can further contribute to the photochemical processes of the atmosphere. The ramifications of photoelectrons in exoplanetary atmospheres has received limited focus so far, mainly on their role in establishing the He population in the upper atmosphere \citep{TaylorA25, GarciaMunoz25a, GarciaMunoz25b}. However, these high-energy electrons can further excite, dissociate and/or ionise the atmospheric gases, therefore increase the formation of disequilibrium products. We investigate this process here to evaluate if the enhanced photochemistry could potentially increase the production of C$_2$H$_4$ and potential of other species that would improve the fits of the MIRI observations. Nevertheless, beyond the implications for K2-18 b, photoelectrons could have a far reaching role for the understanding of disequilibrium chemistry at different exoplanetary atmospheres.

\begin{figure}[!h]
\centering
\includegraphics[scale=0.5]{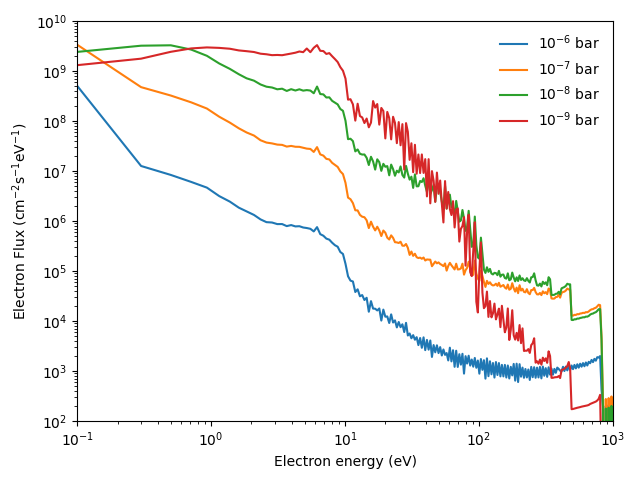}
\caption{Energy spectrum of secondary electrons at different pressure levels in the atmosphere of K2-18 b.}\label{secondaries}
\end{figure}

\begin{figure}[!h]
\centering
\includegraphics[scale=0.5]{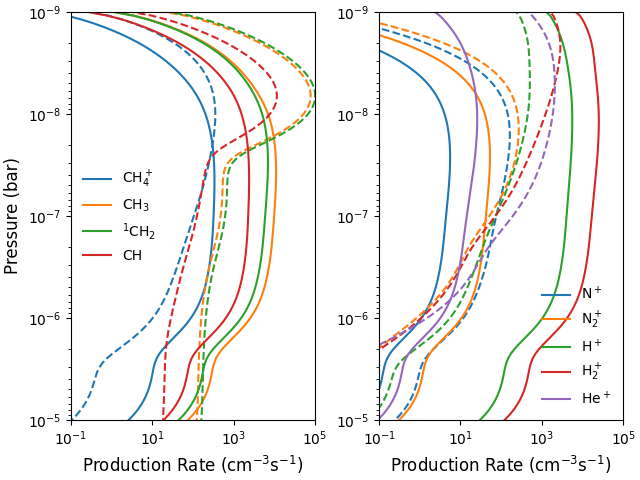}
\caption{Comparison of production rates for different products of dissociation or ionisation by photons (dashed lines) and photoelectrons (solid lines) in the atmosphere of K2-18 b.}\label{fragments}
\end{figure}

\begin{figure*}
\centering
\includegraphics[scale=0.48]{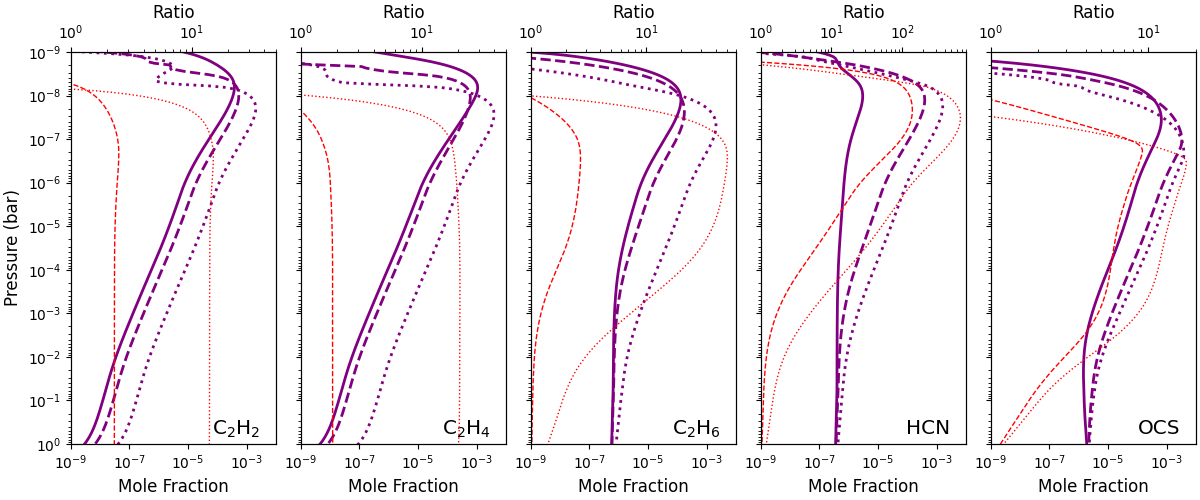}
\caption{Comparison of mole fractions of various photochemical products among the nominal case (solid purple lines), the enhanced photochemistry with photoelectrons (dashed purple lines) and the overall effect of increased UV flux of the star (dotted purples lines). The red lines present the ratios of the corresponding line style cases to the nominal case.}\label{photochem}
\end{figure*}

To investigate the impact of such processes on the atmospheric composition of K2-18 b we used an electron energy deposition model developed for Titan's atmosphere, \citep{Lavvas11}. Titan's atmosphere is mainly composed of N$_2$ and CH$_4$, and we expanded this simulation with the inclusion of electron impact cross sections for H$_2$, He and H. This combination of cross sections covers the main gaseous abundances of the atmosphere that will be responsible for the dominant generation of photoelectrons and their energy degradation. The code uses as input the stellar energy flux at high energies ($\lambda$$<$300 nm) and calculates the photoionization rates of the atmospheric compounds at different altitudes in the atmosphere. The energy of the produced photoelectrons depends on the difference between the energy of the ionising photon and the ionisation potential for the specific photoionisation or dissociative-photoionisation pathway of each target gas species. The results of this calculation demonstrate that most of the photoelectrons generated originate from the ionisation of H$_2$ and He, while CH$_4$ followed by N$_2$ have overall a secondary role due to their lower abundance (Fig.~\ref{photoelectrons}). Atomic hydrogen, due to its rather small abundance in this atmosphere has a minor contribution.  However, there is a clear influence on the pressure range where each species contributes to the photoelectron production: CH$_4$ and N$_2$ affected by diffusive separation have decreasing abundances at pressures lower than $\sim$10 nbar, which constraints their dominant photoelectron production to higher pressures. Instead H$_2$ and He have significant abundances and allow the formation of photoelectrons at lower pressures. For each gas species, the photoelectron production demonstrates an increasing production towards higher pressures as the photoelectron energy increases. This characteristic arises from the decreasing photoabsorption cross section of gas species towards high photon energies that are required to generate such photoelectrons, i.e. photons at these energies penetrate deeper in the atmosphere and allow to generate photoelectrons at lower altitudes than those of neutral photodissociation as we will see below. The combined photoelectron production reveals that high energy photoelectrons able to further ionise and dissociate the main gas species are generated down to $\sim$10$^{-5}$ bar. 

The generated photoelectrons are then locally degraded in energy through collisions with gases and thermal electrons. These collisions can result to the excitation, dissociation or ionisation of the gases with all these processes shaping the resulting energy distribution of the produced secondary electrons. The calculated spectra (Fig.~\ref{secondaries}) reveal an increasing flux of electrons with decreasing energy, while the structure observed in the spectra depends on the relative contributions of photoelectron production at each pressure level and their energy loss from the collisions with different gases. For example at 10$^{-9}$ bar the electron energy spectrum shows a steep slope between 10 and 1000 eV since most of the photoelectrons produced are in the 10-100 eV range, while at higher pressures significant production of photoelectrons at higher energies results {\color{\clr}in} a flatter energy distribution of the electrons in this energy range. The shape and magnitude of the electron energy spectrum controls the dissociation/ionisation of the gaseous species depending on their electron impact cross sections \citep[see e.g. figs. 14 and 15 in][]{Lavvas11}.

The overall effect of interest here is the additional fragmentation of the parent gas species that would enhance the disequilibrium chemistry of the atmosphere. These fragments include CH$_4^+$, CH$_3$, CH, H, H$^+$, H$_2^+$, He$^+$, N, N($^2$D), N$^+$ and N$_2^+$. Comparing the production of these fragments from photon and electron impact processes (Fig.~\ref{fragments}) we can clearly notice that for the fragments induced from CH$_4$, photons provide the major contribution at pressures near 10$^{-8}$ bar, while photoelectrons dominate the fragmentation processes at higher pressures. For N$_2$, as the photons allowing ionisation ($\lambda$$<$799 \AA) are readily absorbed by the more abundant H$_2$, photoelectrons have a dominant contribution to the production of N$^+$ and N$_2^+$. The production rates of these fragments are then input in the disequilibrium chemistry simulations to evaluate their role in the resulting abundance of various photochemical products. 

\begin{figure*}
\centering
\includegraphics[scale=0.45]{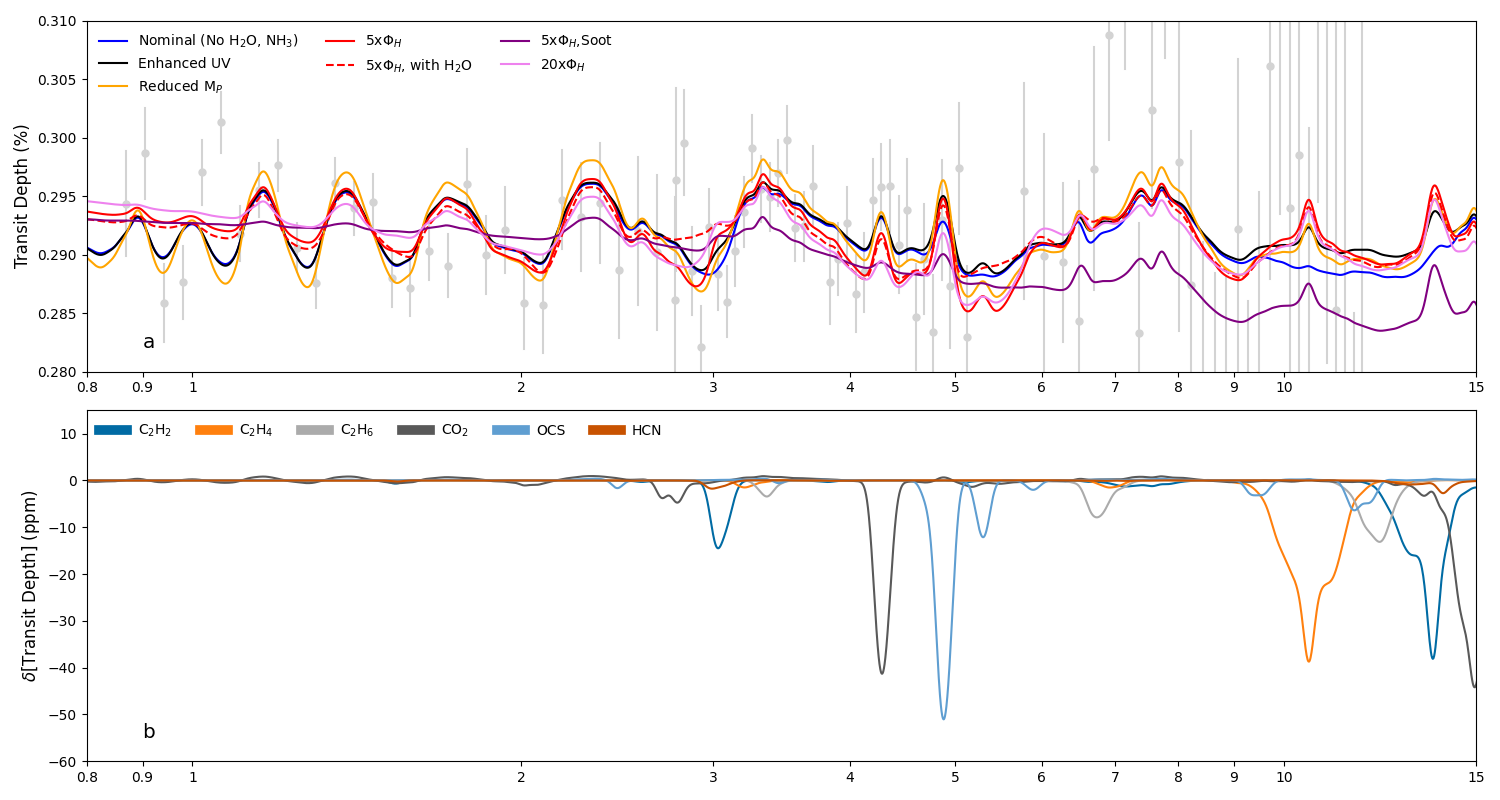}
\caption{Sensitivity tests on transit spectrum. Panel a compares the nominal spectrum for the 200$\times$solar metallicity case and T$_{int}$=60K case (blue line) with different scenarios discussed in the text and related to the planet mass, the haze mass flux and its optical properties. Panel b presents the differential transit signature of different gases for the enhanced UV scenario.}\label{transit_2}
\end{figure*}

Our simulations reveal that the role of photoelectrons is significant, affecting the overall photochemistry of the upper atmosphere (compare solid and dashed lines in Fig.~\ref{photochem}). The increased fragmentation of CH$_4$ increases the abundance of pure hydrocarbons such as C$_2$H$_2$, C$_2$H$_4$ and C$_2$H$_6$ by factors of $\sim$2 relative to the simulations without photoelectrons. The impact on the nitrogen chemistry is far greater since the enhanced fragmentation of N$_2$ leads to an enhancement of nitriles, e.g. the HCN mole fraction increases by factors up to 150. The ramifications of the enhanced photochemistry propagate as well to species not directly affected by photoelectrons in our simulations, for example OCS also presents increased mole fractions by factors up to $\sim$10. However, when included in the transit simulation, the enhanced abundances do not significantly affect the spectrum, apart from an increase in the transit depth for OCS (Fig.~\ref{transit_2}). The increase in HCN occurs at pressures that are not probe{\color{\clr}d} in transit while the increase of C$_2$H$_4$ is barely visible in the simulated spectrum.

As described earlier our nominal stellar flux for K2-18 is based on the flux of the similar GJ 176 star, with the former however being a much young star than the latter. Therefore the high energy output of K2-18 could be higher and we thus further explore the ramifications of an enhanced UV stellar flux. We evaluate the difference in the atmospheric composition for an increased stellar output by a factor of 4 at wavelengths below 300 nm, as indicated by the measured Ly-$\alpha$ flux of K2-18 b \citep{Santos20}, taking into consideration both the role of photons and photoelectrons. Following the same procedure as above, our simulated abundances indicate a significantly larger enhancement on the mole fractions of the photochemical products discussed earlier (see dotted lines in Fig.~\ref{photochem}). The combined effect of photon and photoelectrons results in maximum increases of  $\sim$15$\times$, 20$\times$, 50$\times$, 700$\times$, 17$\times$ higher C$_2$H$_2$, C$_2$H$_4$, C$_2$H$_6$, HCN and OCS mole fractions, respectively. The enhancement of the mole fractions at the pressure region probed in transit is sufficient to have a clear signature of C$_2$H$_4$ on the simulated transit spectrum (compare blue and black lines in Fig.~\ref{transit_2}), while the contributions of C$_2$H$_2$ and C$_2$H$_6$ are also becoming visible. Despite its large enhancement, HCN is not affecting the transit spectrum significantly. The increase in transit depth at MIRI wavelengths due to the enhanced photochemistry is smaller than the uncertainty of the observations, while the amplitude of the spectral features remains small compared to the observations. Nevertheless, irrespective of the implications of this enhanced photochemistry on the transit spectra the above results demonstrate that photoelectrons have an important chemical impact and merit further attention by the exoplanet community.

\section{JWST/MIRI transit amplitude}

Regardless of the large uncertainty of the JWST/MIRI observations, we explore here processes that could potentially increase the amplitude of the simulated transit depth with the goal to understand how sensitive our simulated spectra are at different conditions. We thus keep the enhanced UV stellar flux scenario and investigate mechanisms that would enlarge the atmospheric scale height at the region probed in transit,  such as a reduced energy redistribution factor, modified haze parameters (mass flux, optical properties) and a reduced planetary mass. We stress that the following tests are not intended as best-fit solutions, but as sensitivity experiments designed to assess how the transit amplitude responds to plausible physical variations.

Thus far our simulations were performed using a planetary mass of M$_P$ = 8.63$\pm$1.35M$_{\odot}$ \citep{Benneke19}, however the latest evaluation with K2 observations provides a slightly reduced mass of 7.2$^{+1.5}_{-1.4}$M$_{\odot}$ \citep{Howard25}. Previous studies have identified the degeneracy between planet mass and atmospheric temperature, which control the atmospheric scale height \citep{Liu25}, and using the lower bound on the planet mass for our simulations provides a larger scale height with evident ramifications on the transit spectrum (orange line in Fig.~\ref{transit_2}): the amplitude between peaks and troughs is now higher in the MIRI region (although still smaller than the observed), while the rest of the spectrum remains consistent with the observations at shorter wavelengths. We thus keep this planet mass value for the following evaluations.

\begin{figure}
\centering
\includegraphics[scale=0.55]{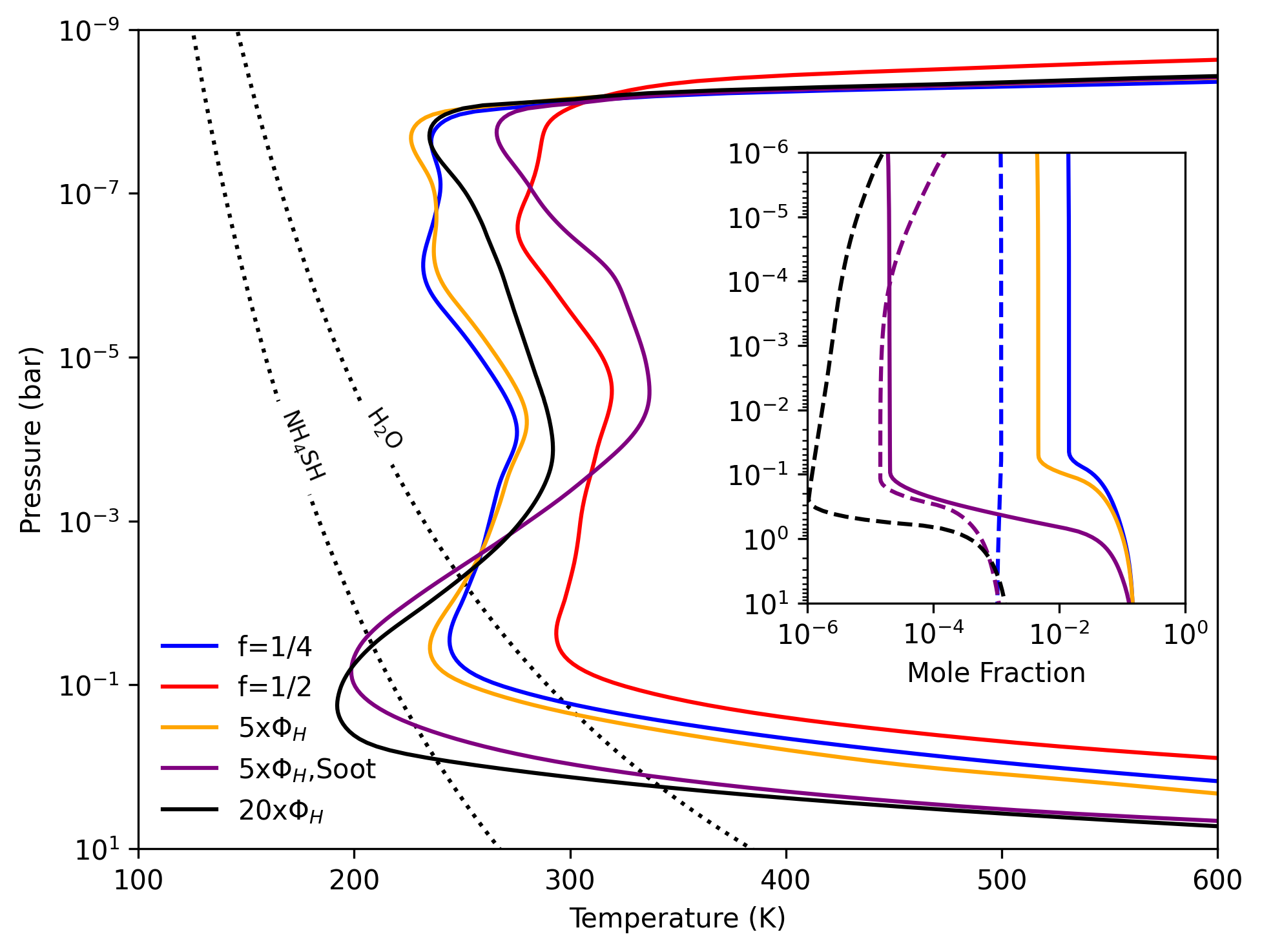}
\caption{Sensitivity results of atmospheric thermal structure on different scenarios for a 200x solar metallicity atmosphere. The blue presents the nominal scenario of complete energy redistribution and $\rm\phi_H$=10$^{-12}$ gcm$^{-2}$s$^{-1}$ with haze optical properties from \cite{He24}. The orange and black lines present a 5$\times\rm\Phi_H$ and 20$\times\rm\Phi_H$ scenarios with the same optical constants, while the purple line demonstrates the impact of a more absorbing compositions such as that of soots \citep{Lavvas17}. The red line corresponds to the nominal $\rm\Phi_H$ and refractive index case, but energy deposition on the day side. The dotted lines present the condensation curves for H$_2$O and NH$_4$SH. The inset presents the impact of each scenario on the H$_2$O (solid lines) and NH$_3$ (dashed lines) mole fractions, with colours corresponding to the temperature cases of the main plots.}\label{tempsense}
\end{figure}

A higher temperature in the probed region would further increase the atmospheric scale height. Our thermal structure simulations until now assumed a complete energy redistribution of the stellar energy in the atmosphere of K2-18 b. Considering that the stellar energy is distributed only on the day side allows for an increase of $\sim$50 K throughout the atmosphere (red line in Fig.~\ref{tempsense}). However such an increase would limit the condensation of H$_2$O, thus increase its contribution to the transit spectrum, contrary to the observations. Therefore, we do not consider such a scenario valid.

Another mechanism to increase the temperature locally in the stratosphere of K2-18 b would be to have more absorbing haze particles. The optical properties from \cite{He24} used to evaluate the interaction of the haze particles with the stellar radiation result in significant scattering but are characterised by weak absorption, thus generate a weak contribution to the atmospheric heating. Assuming the slightly more absorbing optical properties of Titan's haze particles \citep{Lavvas10}, increases the atmospheric temperature by a few kelvin but provides an overall similar picture (not shown). Keeping the \cite{He24} refractive index and raising the haze mass flux above the conservative $\rm\Phi_H$=10$^{-12}$ g cm$^{-2}$s$^{-1}$ scenario by a factor of 5, slightly warms the atmosphere {\color{\clr}at pressures lower than} $\sim$10 mbar and cools it {\color{\clr}at higher pressures} since less photons arrive there compared to the nominal haze case (orange line in Fig.~\ref{tempsense}). The resulting transit spectrum reveals an enhanced transit depth at short wavelengths with smaller variability due to the increased haze opacity (red line in Fig.~\ref{transit_2}). The reduced temperature in the lower atmosphere enhances the H$_2$O condensation further reducing its gaseous abundance by a factor of $\sim$3, (see inset in Fig.~\ref{tempsense}), which however is not sufficient to completely remove the H$_2$O signature from the spectrum (compare red solid and dashed lines in Fig.~\ref{tempsense}). The reduction of H$_2$O also enhances the photolysis of CH$_4$ that drives an increased C$_2$H$_4$ production that raises its signature on the transit spectrum. Therefore, we conclude that a stronger haze contribution would further improve the agreement of the simulated spectra with the observations.

If, in addition to the above we considered a more absorbing particle composition such as that of soot \citep{Lavvas17}, the atmosphere would be substantially heated at pressure{\color{\clr}s lower than} 1 mbar and further cooled {\color{\clr}at deeper levels} (purple line in Fig.~\ref{tempsense}). Under such conditions the amplitude of the molecular features would be further increased, but our simulations demonstrate that the haze contribution dominates the atmospheric opacity and provides eventually a far more muted spectrum (gray line in Fig.~\ref{transit_2}). Certainly the presence of soot type composition would be surprising for the cool atmosphere of K2-18 b, but this test suggests that if more/stronger absorption bands were present in the photochemical hazes of such atmospheres, they would also impose conditions closer to the transit observations. We discuss below possible ways such a modification could occur.

\section{The lack of NH$_3$}

The drop in the atmospheric temperature below 10 mbar when haze opacity is increased could have further ramifications for the lack of NH$_3$ in K2-18 b's transit spectrum. Our simulations indicate that the reduced temperature conditions could allow the formation of ammonium hydrosulphide, NH$_4$SH (Fig.~\ref{tempsense}). This condensate has been long speculated to form in the giant planets of the solar system \citep{Weidenschilling73, West04}, while it is an anticipated ice in the interstellar medium \citep{Slavicinska25}. Its formation is predicted from thermochemical equilibrium arguments and its condensation curve at the high metallicity conditions explored here indicate that it should form near the temperature minimum at 100 mbar (Fig.~\ref{tempsense}). We evaluated the condensation curve using the methodology of \cite{Visscher06}, for the formation through reaction:
\begin{equation}
\rm NH_3 + H_2S \rightarrow NH_4SH(s),
\end{equation}
but for high metallicity conditions (200xSolar) under which NH$_3$ abundance is 1/10th of that of N$_2$ and with the equilibrium constant of \cite{Lewis69}. For these conditions the condensation curve follows the relationship:
\begin{equation}
\rm \frac{10^4}{T} = 49.309 - 4.25 (logP + log[Fe/H] - 1)
\end{equation}
with P the pressure in bar, T the temperature in Kelvin and Fe/H the atmospheric metallicity.
The formation of this cloud would result in a significant loss of NH$_3$ while the abundance of H$_2$S would be reduced by a few percent. We explored the formation of such a cloud and its impact on the NH$_3$ abundance in the atmosphere of K2-18b, considering a surface tension of 100 dyne/cm in lack of explicit information. Our results demonstrate that for the highly absorbing haze scenario explored above, the NH$_4$SH cloud reduces the NH$_3$ abundance by a factor of $\sim$100 (see inset in Fig.~\ref{tempsense}), thus removing naturally its signature from the simulated transit spectrum, in agreement with the observations. Moreover, the NH$_4$SH cloud has a similar upper boundary to the H$_2$O cloud with particles reaching maximum radii of $\sim$100 $\mu$m, therefore its formation has a negligible impact on the transit spectrum. 

Despite the appeal of this simple solution for the interpretation of the apparent lack of NH$_3$ in the transit spectrum of K2-18 b, the highly absorbing (soot refractive index) haze scenario that allowed for the strong temperature minimum near 10 mbar is not anticipated. However, an increased haze mass flux leads to a locally colder atmosphere in the region of interest, as we saw above. We thus explored how much more reflecting haze (with the nominal refractive index considered) would be required to have a sufficient cooling of the atmospheric temperature to allow the NH$_4$SH formation. Our simulations indicate that a mass flux of 2$\times$10$^{-11}$ gcm$^{-2}$s$^{-1}$, i.e. 20$\times\rm\Phi_H$, would bring the atmospheric temperature well below the NH$_4$SH condensation curve (black line in Fig.~\ref{tempsense}) and result practically in the removal  of the gaseous NH$_3$. At the same time the resulting transit spectrum remains consistent with the observations (violet spectrum in Fig.~\ref{transit_2}). Is however, such a high haze mass flux possible?

\section{Discussion}

\begin{figure}
\centering
\includegraphics[scale=0.53]{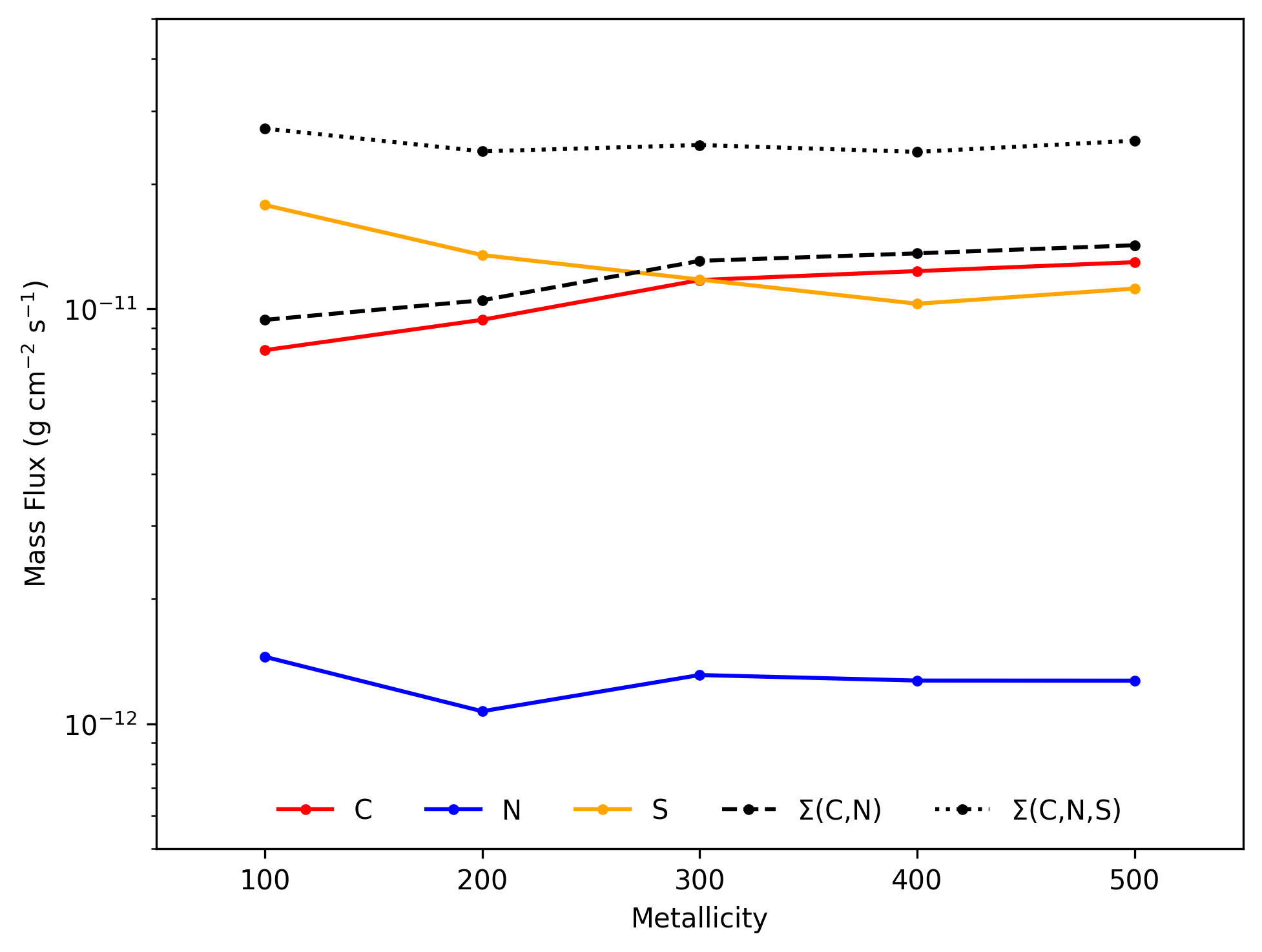}
\caption{Mass fluxes generated from the photolysis of precursor gases in the upper atmosphere (p$<$10$\mu$bar) of K2-18 b. Reb, blue and orange lines correspond to precursors of C, N and S composition, respectively. The black dashed line sums contributions from C and N precursors and the black dotted line includes also the sulphur contribution. The results correspond to the nominal UV flux.}\label{photorates}
\end{figure}

Our sensitivity tests described above demonstrate that a higher haze mass flux would improve the overall agreement of the forward model results with the transit observations. The nominal mass flux scenario ($\rm\Phi_H$=10$^{-12}$ gcm$^{-2}$s$^{-1}$) corresponds to $\sim$10$\%$ of the mass flux generated from the C- and N-based  precursors (Fig.~\ref{photorates}) under nominal conditions, while for the more likely enhanced disequilibrium chemistry scenario the precursor photolysis mass flux would increase by $\sim$4$\times$. This picture applies to all metallicity cases considered in our evaluation. Therefore it is justified to assume a more efficient photochemical haze production.

Furthermore, while the contributions from C and N containing precursors are significant, contributions from sulphur containing species have a comparable mass flux and they could partake in the formation of photochemical hazes at the simulated conditions (Fig.~\ref{photorates}). Laboratory experiments on the formation of photochemical hazes demonstrate that the inclusion of H$_2$S in the gas mixture used for the production of the particles, rapidly increases the haze formation yield \citep{Yu20,Vuitton21,Maratrat25}. What is certainly missing and critical for the role of haze in K2-18 b and similar sub-Neptunian temperate atmospheres are the optical properties of such organosulphuric hazes and specifically if inclusion of sulphur makes the resulting particles more absorbing. The published experiments indicate that the sulphur tholins have a grey hue compared to the orange tholins of C/N composition, indicating that there is an overall modification of the material absorptivity at visible wavelengths. Similar experiments on the optical properties of the organic residue formed during the irradiation of organic ice mixtures also demonstrate the formation of new absorption bands at 1.8 $\mu$m, 4.9$\mu$m and at longer wavelengths when H$_2$S is included in the mixture, while bands attributed to C and N bonds are reduced in magnitude \citep{Mahjoub21, Mahjoub23}. Although these characteristics are not sufficient to justify a higher heating role of the hazes in the atmosphere of K2-18 b, they do point in the right direction and highlight the need for the optical properties of organosulphuric particles.

In view of the above, the haze mass flux of 20$\times$$\rm\Phi_H$ used to cool enough the atmosphere for NH$_4$SH formation to occur, is consistent with the potential of the atmospheric photochemistry to generate such particles, particularly when sulphur precursors are included in the estimation of photolysis mass fluxes and when the enhanced UV output of K2-18 is taken in consideration. This upper limit of 20$\times$$\rm\Phi_H$ could be further reduced to values that would allow for a gaseous NH$_3$ abundance just below the detection limit, while a more absorbing refractive index for the haze particles would further cool locally the atmosphere thus reduce the required haze mass flux to achieve the same condensation conditions. Moreover, including the impact of the clouds on the thermal structure would aid in the same direction. We thus suggest that a strong photochemical haze production in K2-18b's atmosphere provides a physically plausible mechanism for suppressing both H$_2$O and NH$_3$ from its observed transit spectrum.

The atmospheric scenario proposed here leads to several observationally testable predictions for K2-18 b or for a similar exoplanet amenable to better characterisation. The presence of C$_2$H$_4$ and the role of enhanced disequilibrium chemistry could be evaluated with additional observations to improve the signal-to-noise ratio of MIRI/LRS observations. The role of hazes could be further evaluated with improved precision transit observations at similar or shorter wavelengths where haze opacity increases, as well as, with phase curve analysis that is more sensitive to scattering by the particles.
Finally, if NH$_3$ is removed primarily through NH$_4$SH condensation, it should remain undetectable even with improved near-infrared sensitivity, whereas a robust detection of NH$_3$ would disfavour strong haze-induced cooling.

\section{Conclusions}

We have investigated the atmosphere of K2-18 b using a self-consistent one-dimensional model that couples radiative energy deposition, disequilibrium chemistry, and haze and cloud microphysics. Treating K2-18 b as a sub-Neptune, our simulations provide a coherent physical framework that reproduces the key characteristics of the observed JWST transmission spectra.

Our primary result is that a high-metallicity H$_2$-rich atmosphere, with metallicities in the range of approximately 200-400$\times$ solar, offers the most consistent explanation of the observations. This conclusion is robust across multiple JWST data reduction pipelines and is only weakly sensitive to assumptions regarding haze mass flux. A moderate intrinsic temperature (T$_{int}$ = 60 K) is favoured, as it enhances the contribution of CO$_2$ to the transmission spectrum in agreement with the data.

The atmospheric composition of K2-18 b is strongly shaped by disequilibrium chemistry. Methane provides the dominant spectral signature, while CO$_2$ and OCS contribute additional features at near- and mid-infrared wavelengths. A marginal contribution from C$_2$H$_4$ near 10 µm is predicted, particularly under enhanced stellar UV conditions, although current MIRI uncertainties limit the detectability of this feature.  Our forward-model results provide a physical interpretation consistent with our retrieval-inferred abundances \citep{Liu25}, while highlighting degeneracies that cannot be resolved without self-consistent modelling. Importantly, our forward-model results reproduce the observed spectra without invoking more complex molecular species. 

Photochemical hazes emerge as a key component of the atmospheric structure. While water clouds form efficiently, their opacity resides too deep in the atmosphere to directly affect the transmission spectrum. Nevertheless, cloud formation critically limits the abundance of gaseous H$_2$O in the observable atmosphere. A sufficiently large haze mass flux, well within the range permitted by photochemical production rates, cools the atmosphere between 10 and 100 mbar and can entirely suppress the H$_2$O signature in transmission.

We also examined the role of photoelectrons, which are commonly neglected in exoplanet photochemical models. Our results show that photoelectrons significantly enhance the production of hydrocarbons and nitrogen-bearing species, particularly HCN, although the impact on current transmission spectra is modest. This suggests that photoelectron processes may become increasingly important as observational precision improves.

Finally, we show that strong haze-induced cooling can allow the condensation of ammonium hydrosulphide (NH$_4$SH), providing a physically plausible explanation for the absence of NH$_3$ in the observed transmission spectrum of K2-18 b. This mechanism depends on uncertainties in haze mass flux and optical properties, which remain poorly constrained and motivate further laboratory and theoretical work. However, it highlights the links among metallicity, photochemistry, haze production, and cloud formation into a unified interpretation of the observations.

In summary, our results support a picture of K2-18 b as a hazy, high-metallicity sub-Neptune whose observable atmosphere is shaped by strong disequilibrium chemistry and haze/cloud processes. The prominent role of hazes, sulphur chemistry, and photoelectrons highlighted here is likely to be relevant for a broader population of temperate exoplanets and underscores the need for self-consistent forward modelling in tandem with retrieval methods.

\begin{acknowledgements}
This work was supported by the Action Thématique Exosystèmes of CNRS/INSU, co-funded by CEA and CNES.
\end{acknowledgements}


%
   \bibliographystyle{aa} 
   \bibliography{refs_K2-18} 

\end{document}